\documentclass[prb,preprint,showpacs,preprintnumbers,amsmath,amssymb]{revtex4}
\usepackage{bm}

\begin{document}
\title{Spin polarization in  biased Rashba-Dresselhaus
two-dimensional electron systems}

\author{P. Kleinert}
\affiliation{Paul-Drude-Intitut f\"ur Festk\"orperelektronik,
Hausvogteiplatz 5-7, 10117 Berlin, Germany}
\author{V.V. Bryksin}
\affiliation{A.F. Ioffe Physical Technical Institute,
Politekhnicheskaya 26, 194021 St. Petersburg, Russia}
\date{\today}

\begin{abstract}
Based on spin-charge coupled drift-diffusion equations, which are
derived from kinetic equations for the spin-density matrix in a
rigorous manner, the electric-field-induced nonequilibrium spin
polarization is treated for a two-dimensional electron gas with
both Rashba and Dresselhaus spin-orbit coupling. Most emphasis is
put on the consideration of the field-mediated spin dynamics for a
model with equal Rashba and Dresselhaus coupling constants, in
which the spin relaxation is strongly suppressed. Weakly damped
electric-field-induced spin excitations are identified, which
remind of space-charge waves in crystals.
\end{abstract}


\maketitle

\section{Introduction}
Spin-dependent transport phenomena are of great interest to both
basic research and device applications. Especially,
semiconductor-based spin electronics has been the subject of
numerous investigations. In this field, spin rather than charge is
exploited for signal processing. The spin-orbit interaction (SOI)
opens the possibility of manipulating the spin of carriers using
purely electrical means. Unfortunately, the very same SOI has the
undesired effect of causing spin relaxation due to precession in a
wave-vector dependent effective magnetic field, which is traced
back to the SOI. For a quantum well grown on a [001] substrate,
the Dyakonov-Perel spin relaxation \cite{Dyakonov} is the most
dominant effect. In general, however, spin relaxation depends on
the details of the band structure and the relevant scattering
mechanisms (see Ref. [\onlinecite{JAP_073702}] and references
therein). For an asymmetric quantum well, bulk-inversion asymmetry
and structure-inversion asymmetry give rise to Dresselhaus and
Rashba spin-orbit contributions to the Hamiltonian, respectively.
The interplay between the linear Rashba and Dresselhaus terms
causes a spin relaxation anisotropy that has been treated in a
number of theoretical
works.~\cite{PRB_15582,JPC_R271,PRL_146801,PRB_165311,Pershin_E}
For a quantum well grown along the [001] direction, the main axes
of the spin-relaxation-time tensor are given by [110] and
[1$\overline{1}$0]. Most interesting is the observation that under
idealized conditions and when the linear Rashba and Dresselhaus
terms have equal strength, the relaxation of spin oriented along
the [110] axis is totally suppressed.~\cite{PRB_15582} In this
particular case, a conserved quantity exists, which hinders spin
randomization.~\cite{PRL_146801,PRB_165311} This remarkable
behavior of the spin-relaxation time lead to the proposal of a
nonballistic spin-FET.~\cite{PRL_146801} Other studies of the
combined Rashba-Dresselhaus model referred to the spin- and
charge-Hall
effect.~\cite{PRB_081312,PRB_085315,PRB_121308,PRB_085344,IJMPB_4937,PRB_155323}

Recently, the field received a fresh impetus by the identification
of an exact SU(2) symmetry of the model, when reaching the
condition of equal Rashba and Dresselhaus coupling
strength.~\cite{PRL_236601,PRB_235322} From a theoretical point of
view, an equivalent system is the Dresselhaus [110] model. The
revealed symmetry gives rise to a massless mode with infinite
lifetime at nonzero wave vector. Qualitative features of the
associated spin pattern have been experimentally confirmed
\cite{PRL_076604W} by optical techniques that probe spin
relaxation rates. Furthermore, it has been predicted
\cite{PRB_241308,PRB_125307} that coherent spatial oscillations of
the spin polarization develop in such a system under appropriate
injection conditions.

In this paper, we extend these interesting studies by treating
spin effects under the influence of an applied in-plane electric
field. For the combined Rashba-Dresselhaus model with a SOI that
is linear in $\bm{k}$, spin-charge coupled drift-diffusion
equations are derived in a systematic manner. Special results are
presented and discussed for the special model with equal Rashba
and Dresselhaus coupling strengths.

\section{Spin-charge coupled drift-diffusion equations}
The effective Hamiltonian of our approach
\begin{equation}
H_0=\frac{\hbar^2\bm{k}^2}{2m}+\alpha(k_y\sigma_x-k_x\sigma_y)+\beta(k_x\sigma_x-k_y\sigma_y)
\label{H0}
\end{equation}
includes the Rashba spin-orbit term, which is due to the inversion
asymmetry of the confining potential of the quantum well. In
addition, there is the Dresselhaus coupling, which is present in
semiconductors lacking bulk inversion symmetry. The model
Hamiltonian refers to a two-dimensional semiconductor
nanostructure grown along the [001] direction. In Eq.~(\ref{H0}),
$\bm{k}$, $m$, and $\sigma_{i}$ ($i=x,y,z$) denote the in-plane
wave vector, the effective electron mass, and the usual Pauli
matrices, respectively. $\alpha$ and $\beta$ are the strengths of
the Rashba and Dresselhaus spin-orbit couplings. Our total
Hamiltonian encompasses also contributions stemming from the
short-range spin-independent elastic scattering on impurities and
the in-plane electric field $\bm{E}=(E_x,E_y,0)$. Its explicit
form together with related kinetic equations for the spin-density
matrix has been published recently.~\cite{SSC_139} It should be
noted that for the combined Rashba-Dresselhaus model, the
field-induced spin polarization depends on the orientation of the
in-plane electric field.~\cite{IJMPB_4937,PRB_155323}

The main quantity for the theoretical analysis of spin-related
phenomena is the spin-density matrix
$f_{\lambda^{\prime}}^{\lambda}(\bm{k},\bm{k}^{\prime},t)$, which
is calculated from quantum-kinetic equations.~\cite{PRB_165313}
From this set of equations, spin-charge coupled drift-diffusion
equations are derived for the physical components
$\overline{f}={\rm Tr}\overline{\widehat{f}}$ and
$\overline{\bm{f}}={\rm Tr}\overline{\bm{\sigma}\widehat{f}}$,
which are integrated over the polar angle of $\bm{k}$ (denoted by
the bar). In the case of weak SOI ($\alpha k_F\tau/\hbar$, $\beta
k_F\tau/\hbar \ll 1$, with $k_F$ and $\tau$ being the Fermi wave
vector and elastic scattering time, respectively), the following
ansatz is justified \cite{PRB_075340}
\begin{equation}
\overline{f}({\bm{k}},{\bm{q}},t)=-F({\bm{q}},t)\frac{d
n(\varepsilon_{\bm{k}})/d\varepsilon_{\bm{k}}}{dn/d\varepsilon_F},\quad
\overline{{\bm{f}}}({\bm{k}},{\bm{q}},t)=-{\bm{F}}({\bm{q}},t)\frac{d
n(\varepsilon_{\bm{k}})/d\varepsilon_{\bm{k}}}{dn/d\varepsilon_F},
\label{ansatz}
\end{equation}
where $n(\varepsilon_{\bm{k}})$ denotes the Fermi function and
$\varepsilon_{\bm{k}}=\hbar^2\bm{k}^2/(2m)$. Furthermore, we
introduced the electron density $n=\int d\varepsilon
\rho(\varepsilon)n(\varepsilon)$, with $\rho(\varepsilon)$ being
the density of states of the two-dimensional electron gas.
Applying this straightforward calculational
scheme~\cite{PRB_075340}, spin-charge coupled drift-diffusion
equations are obtained, which read in spatial coordinates $r_j$
\begin{equation}
\frac{\partial F_i}{\partial t}+\frac{\partial J_{ij}}{\partial
r_j}+M_{ij}F_j=\frac{2m\tau}{\hbar^3\tau_s}L_i F, \label{b1}
\end{equation}
with the expression for the spin flux
\begin{equation}
J_{ij}=\left(\mu E_{j}-D\frac{\partial}{\partial r_{j}}
\right)F_{i}+\frac{4Dm}{\hbar^2}\varepsilon_{ikl}Q_{kj}F_{l}
-\frac{4mD\tau}{\hbar^3}(\alpha^2-\beta^2)
\left[\frac{2m}{\hbar^2}Q_{ij}-\varepsilon_{ijk}\frac{\mu
E_{k}}{2D} \right]F. \label{JJ}
\end{equation}
The diffusion coefficient $D$ and the mobility $\mu$ satisfy the
Einstein relation $\mu=(eD/n)dn/d\varepsilon_F$. Other quantities,
which enter Eqs.~(\ref{b1}) and (\ref{JJ}), are defined by
\begin{equation}
{\bm{L}}=\frac{2m}{\hbar^2}\tau_s\left(-(\alpha\mu E_y+\beta\mu
E_x),(\alpha\mu E_x+\beta\mu E_y),0 \right),\quad
\frac{1}{\tau_s}=4D\frac{m^2}{\hbar^4}(\alpha^2 +\beta^2),
\end{equation}
\begin{equation}
M_{ij}=\frac{A_{ij}}{\tau_s}+\frac{1}{\tau_s}\varepsilon_{ikj}L_k
+\frac{2\alpha\beta}{(\alpha^2+\beta^2)\tau_s}S_{ij},
\end{equation}
\begin{equation}
\overleftrightarrow{A}=\left(%
\begin{array}{ccc}
  1 & 0 & 0 \\
  0 & 1 & 0 \\
  0 & 0 & 2 \\
\end{array}%
\right),\quad \overleftrightarrow{Q}=\left(%
\begin{array}{ccc}
  \beta & \alpha & 0 \\
  -\alpha & -\beta & 0 \\
  0 & 0 & 0 \\
\end{array}%
\right),\quad \overleftrightarrow{S}=\left(%
\begin{array}{ccc}
  0 & 1 & 0 \\
  1 & 0 & 0 \\
  0 & 0 & 0 \\
\end{array}%
\right).
\end{equation}
$\varepsilon_{ijk}$ is the totally antisymmetric tensor in three
dimensions. The spin flux $J_{ij}$, which gives the $i$'th
component of the particle flow with spin polarization along the
axis $j$ ($i,j=x,y,z$), has the general form that is in accordance
with symmetry requirements.~\cite{PRB_041308} Neglecting the
spin-charge coupling and the influence of the electric field, we
obtain a result
\begin{equation}
\frac{\partial F_i}{\partial
t}=-\frac{m}{\hbar^2}Q_{jl}\varepsilon_{jik}J_{kl},\quad
J_{kl}=\frac{4Dm}{\hbar^2}\varepsilon_{kmn}Q_{ml}F_n,
\end{equation}
in which the spin flux $J_{kl}$ appears as a source of
nonequilibrium spin polarization.~\cite{PRB_041308,ssc_559} The
set of basic equations (\ref{b1}) and (\ref{JJ}), which restore
published results \cite{PRL_236601} in the case of vanishing
electric field, are solved and discussed in the Fourier space with
respect to spatial coordinates. For the sake of a better
readability of the paper, these equations are summarized in the
Appendix.

\section{Results and discussion}
A number of quite interesting electric-field effects on the spin
polarization occurring in systems with Rashba and Dresselhaus SOI
can be studied on the basis of the coupled spin-charge
drift-diffusion Eqs.~(\ref{b1}) and (\ref{JJ}) [or (\ref{A1}) and
(\ref{A2})]. To keep the presentation transparent, we restrict
ourselves to spatially infinite systems by omitting any boundary
effects.

\subsection{Charge current}
The current density of charge carriers is defined by the time
derivative of the dipole moment, which can be expressed by
\begin{equation}
\bm{j}(t)=-ie\nabla_{\bm{\kappa}}\frac{\partial}{\partial
t}F(\bm{\kappa},t)\mid_{\bm{\kappa}=\bm{0}}.
\end{equation}
As we are mainly interested in the charge transport along the
[110] and [1$\overline{1}$0] directions, the spatial coordinates
are rotated by introducing new wave vectors
$\kappa_{\pm}=(\kappa_x\pm \kappa_y)/\sqrt{2}$. In these
coordinates, the charge current along the $\kappa_{+}$ direction
is given by
\begin{equation}
j_{+}(t)=-ie\frac{\partial^2}{\partial t\partial\kappa_{+}}
F(\kappa_{+},\kappa_{-}=0,t)\mid_{\kappa_{+}=0}.
\end{equation}
Taking into account Eq.~(\ref{A1}), we immediately obtain for the
Laplace transformed current density
\begin{equation}
j_{+}(s)=e\mu
E_{+}\frac{F}{s}+\frac{e}{\hbar}(\alpha+\beta)F_{-}(s)
-\frac{2me\tau}{\hbar^3}(\alpha^2-\beta^2)\mu
E_{-}F_z(s),\label{current}
\end{equation}
where $E_{\pm}=(E_x+E_y)/\sqrt{2}$ and $F_{\pm}=(F_x\pm
F_y)/\sqrt{2}$. The components of the density matrix, which enter
this equation, refer to a spatially homogeneous system. The
related ($\bm{\kappa}=\bm{0}$) drift-diffusion Eqs.~(\ref{A1}) and
(\ref{A2}) are easily solved. Inserting the solution into
Eq.~(\ref{current}) and taking into account only contributions
linear to the electric field, we get
\begin{equation}
\displaystyle j_{+}(\omega)=e\mu
E_{+}F\biggl\{1+\frac{\alpha^2\tau}{\hbar^2D}
\frac{1-\gamma^2}{i\omega\tau_s-{(1-\gamma)^2}/({1+\gamma^2})}
\biggl\},\label{current1}
\end{equation}
with $\gamma=\beta/\alpha$. As the current is aligned along the
electric field direction, there is no Hall current, the appearance
of which would require the treatment of the nonlinear field term
in Eq.~(\ref{current}), which is of higher order in the spin-orbit
coupling constant. In Eq.~(\ref{current1}), the pure charge
current $e\mu E_{+}F$ is complemented by a spin contribution,
which exhibits a Drude-like frequency dependence governed by the
spin-relaxation time $\tau_s$ that may be much larger than the
elastic scattering time $\tau$. Consequently, the frequency
dispersion of the spin-mediated current contribution can set in at
lower frequencies as in the ordinary Drude formula. The
spin-induced current contribution disappears, when the Rashba and
Dresselhaus coupling constants are equal ($\alpha=\beta$). For the
Rashba model ($\beta=0$), Eq.~(\ref{current1}) resembles the
results derived previously for small polarons.~\cite{PRB_235302}

\subsection{Out of plane spin polarization for $\bm{\alpha=\beta}$}
As another application of the spin-charge drift-diffusion
Eqs.~(\ref{A1}) and (\ref{A2}), we treat the evolution of an
initial spin lattice produced at $t=0$ along the $\kappa_{+}$
direction
\begin{equation}
f_{z0}(\kappa_{+})=\frac{1}{2}f_{z0}\delta(\kappa_{+}-\kappa_0)+\frac{1}{2}
f_{z0}^{*}\delta(\kappa_{+}+\kappa_0),\quad f_{z0}=\mid f_{z0}\mid
e^{i\varphi},
\end{equation}
with $\kappa_0$ being a given wave vector. Restricting ourselves
to the special case $\alpha=\beta$, we obtain the exact solution
\begin{equation}
F_z(\kappa_{+},s)=f_{z0}(\kappa_{+})\frac{\sigma}{\sigma^2+2\Omega_{\kappa_{+}}^2},
\label{ee}
\end{equation}
with the effective Laplace variable
\begin{equation}
\sigma=s-i\mu E_{+}\kappa_{+}+D\kappa_{+}^2+2/\tau_s,
\end{equation}
and the frequency
\begin{equation}
\Omega_{\kappa_{+}}=2\sqrt{2}\frac{m\alpha}{\hbar^2}(\mu
E_{+}+2iD\kappa_{+}).
\end{equation}
The inverse Laplace and Fourier transformations of Eq.~(\ref{ee})
are easily calculated and we obtain for the asymptotic dynamics at
large times
\begin{equation}
F_z(r_{+},t)=\frac{\mid f_{z0}\mid}{2}\exp\left[-D(\kappa_0-2K)^2t
\right] \cos\left[\kappa_0r_{+}+\varphi+\mu E_{+}(\kappa_0-2K)t
\right].
\end{equation}
For $\kappa_0=2K$, the original static spin lattice survives and
retains its shape in the steady state. Under this excitation
condition, a long lived spin pattern is expected to occur. If
$\kappa_0$ slightly deviates from $2K$, the electric field drives
a propagating spin wave, the amplitude of which diminishes with
time exponentially. The frequency of this wave is given by $\mu
E_{+}(\kappa_0-2K)$.

\subsection{Field-induced spin accumulation and Hanle effect}
In this subsection, the electric field-induced out-of-plane spin
accumulation is studied for a different initial spin preparation.
First, we focus on the contribution, which appears in a
homogeneous electron gas ($\bm{\kappa}=\bm{0}$). From
Eqs.~(\ref{AA1}) to (\ref{AA4}) given in the Appendix, we obtain
the steady-state solution
\begin{equation}
F_z=-4\alpha\beta\frac{m\tau}{D\hbar^3}\frac{\mu^2(E_x^2-E_y^2)}{2/\tau_s+(\mu\bm{E})^2/D}F,
\label{e44}
\end{equation}
which applies for a semiconductor with different Rashba and
Dresselhaus SOI constants ($\alpha\ne\beta$). This field-mediated
spin accumulation depends on the orientation of the electric field
within the plane and has the character of a second-order field
effect. A spin accumulation, which is proportional to the electric
field, occurs only in the plane.~\cite{IJMPB_4937} The quadratic
field effect in Eq.~(\ref{e44}) disappears for the pure Rashba
($\beta=0$) and Dresselhaus ($\alpha=0$) system as well as under
the condition $|E_x| =|E_y|$. For sufficiently high electric
fields [$2/\tau_s\ll (\mu\bm{E})^2/D$], the out of plane spin
polarization approaches the constant field-independent value
$F_z=-4\alpha\beta m\tau \cos(2\varphi)F/\hbar^3$, with $\varphi$
being the polar angle of the electric field.

In addition to this field contribution of a homogeneous system, we
study the response to a permanent harmonic spin generation of the
form
\begin{equation}
G(r_{+},t)=Ge^{i\omega t+i\kappa_{+}r_{+}}.
\end{equation}
Disregarding the spin-charge coupling, the drift-diffusion
Eqs.~(\ref{AA2}) to (\ref{AA4}) are analytically solved by
\begin{equation}
F_z(\kappa_{+},\omega)=G\frac{(i\Omega+2/\tau_{s+})(i\Omega+2/\tau_{s-})}
{(i\Omega+\frac{2}{\tau_{s+}})(i\Omega+\frac{2}{\tau_{s-}})
(i\Omega+\frac{2}{\tau_s})+T_{+}^2(i\Omega+\frac{2}{\tau_{s-}})
+T_{-}^2(i\Omega+\frac{2}{\tau_{s+}})},\label{disa}
\end{equation}
where the short-hand notations
\begin{equation}
i\Omega=i\omega-i\mu\kappa_{+}E_{+}+D\kappa_{+}^2,\,
T_{+}=2K_{+}(\mu E_{+}+2iD\kappa_{+}),\, T_{-}=2K_{-}\mu E_{-},\,
K_{\pm}=(\alpha\pm\beta)\frac{m}{\hbar^2}
\end{equation}
were used. This solution becomes more transparent for the special
case $\alpha =\beta$, where we obtain
\begin{equation}
F_z(\kappa_{+},\omega)=G\frac{i\omega-i\mu\kappa_{+}E_{+}+D\kappa_{+}^2+2/\tau_s}
{N_{+}N_{-}},\, N_{\pm}=i\omega-i\mu E_{+}(\kappa_{+}\pm
2K)+D(\kappa_{+}\pm 2K)^2 .\label{disb}
\end{equation}
The dispersion relation of eigenmodes $\omega=\mu
E_{+}(\kappa_{+}-2K)$ is derived from the denominator $N_{-}$.
This mode has the character of free carrier oscillations
complemented by a spin part that gives rise to a soft mode at
$\kappa_{+}=2K$.

Further information about the solution in Eq.~(\ref{disa}) is
obtained for $\alpha\approx\beta$ and $\Omega\ll 1/\tau_s$. Two
limits can be distinguished in this case. Under the condition
$\Omega\ll 1/\tau_{s-}$, which applies to the steady state,
Eq.~(\ref{disa}) simplifies to
\begin{equation}
F_z=\frac{G}{2/\tau_s+(\mu\bm{E})^2/D},
\end{equation}
which can be interpreted as the electric-field analogy of the
Hanle effect.~\cite{Kalevich,PRB_075340} Another result is
obtained in the opposite case $1/\tau_{s-}\ll\Omega$, when the
out-of-plane spin polarization depends on the orientation of the
electric field
\begin{equation}
F_z=\frac{G}{2/\tau_s+\left[\mu(E_x+E_y)\right]^2/(2D)}.
\end{equation}
This solution dictates the behavior of the spin polarization in
the limit $\alpha\rightarrow\beta$. Both results agree for the
special field configuration $E_x=E_y=E/\sqrt{2}$.

\subsection{Spin remagnetization waves for $\bm{\alpha =\beta}$}
Finally, we treat the relaxation of an initial homogeneous spin
moment $\bm{F}_0$. In this case, we use $\bm{\kappa}=\bm{0}$ in
our basic equations so that spin and charge degrees of freedom
decouple from each other. Most interesting results are expected,
when the Rashba and Dresselhaus SOI couplings coincide
($\alpha=\beta$). The set of Eqs.~(\ref{AA2}) to (\ref{AA4}) is
easily solved for the Laplace-transformed functions. The result
\begin{equation}
F_x(t)=\frac{F_{x0}-F_{y0}}{2}+e^{-2t/\tau_s}\biggl\{\frac{F_{x0}+F_{y0}}{2}
\cos(\Omega_E t)-\frac{F_{z0}}{\sqrt{2}}\sin(\Omega_E t) \biggl\},
\end{equation}
\begin{equation}
F_y(t)=-\frac{F_{x0}-F_{y0}}{2}+e^{-2t/\tau_s}\biggl\{\frac{F_{x0}+F_{y0}}{2}
\cos(\Omega_E t)-\frac{F_{z0}}{\sqrt{2}}\sin(\Omega_E t) \biggl\},
\end{equation}
\begin{equation}
F_z(t)=e^{-2t/\tau_s}\biggl\{\frac{F_{x0}+F_{y0}}{\sqrt{2}}
\sin(\Omega_E t)+F_{z0}\cos(\Omega_E t) \biggl\},
\end{equation}
describes spin rotations with the frequency
$\Omega_E=\sqrt{2}K\mu(E_x+E_y)$. Such rotations of the magnetic
moment were studied previously both for hopping of small polarons
\cite{PRB_205327,PRB_235302} and for extended states in a
two-dimensional electron gas.~\cite{PRB_205202,PRB_235318} In
analogy to space-charge waves, these eigenmodes are called
spin-remagnetization waves. Typically, the amplitude of these
excitations exponentially decrease with increasing time. It is a
peculiarity of the special Rashba-Dresselhaus model that the
in-plane magnetic moment is conserved:
$F_x(t)-F_y(t)=F_{x0}-F_{y0}$. This observation provides a further
example for the occurrence of undamped spin excitations, whenever
the coupling constants $\alpha$ and $\beta$ are equal.

\section{Summary}
Both for the Rashba-Dresselhaus model with equal coupling
constants and for the Dresselhaus [110] model it has been recently
realized that the relaxation of spin oriented in the [110] axis is
totally suppressed. As the spin lifetime becomes arbitrarily long
within these models, the existence of a persistent spin grating
has been predicted. Recent experiments
\cite{PRL_4196,PRL_076604W,PRL_036603} indeed revealed evidence
supporting this interesting peculiarity of slow spin relaxation
rates. The extreme suppression of spin relaxation in the special
theoretical models for semiconductor nanostructures with SOI is
due to a spin symmetry that gives rise to a soft mode at $\kappa
=2K$. This weakly damped eigenmode leads also to interesting
phenomena in a biased spin-orbit coupled system that was studied
in this paper. Based on rigorous spin-charge coupled
drift-diffusion equations for the components of the spin-density
matrix, a number of field-mediated spin effects were considered:

(i) A Drude-like spin-induced contribution was identified in the
longitudinal charge current that disappears in the special
Rashba-Dresselhaus model with equal coupling strengths. The
frequency position and the width of the resonance are determined
by the spin-relaxation time and the coupling constants $\alpha$
and $\beta$.

(ii) A persistent spin pattern created at $\kappa=2K$ is not
destroyed by the electric field. However, under the condition of
slight-off resonance, the field forces the pattern to move.

(iii) In a homogeneous electron gas with Rashba and Dresselhaus
SOI, an out-of-plane spin polarization develops, which has a
nonlinear field character. In addition, long-lived remagnetization
waves can be excited, whose frequency is given by $\omega=\mu
E_{+}(\kappa_{+}-2K)$. These eigenmodes can be studied in a
similar manner as space-charge waves in crystals. In addition, the
electric-field analogy of the Hanle effect changes its character
at $\alpha\approx\beta$.

(iv) The in-plane spin polarization of the special
Rashba-Dresselhaus model with $\alpha=\beta$ is conserved and the
spins rotate due to the electric field with the frequency
$\Omega_E=\sqrt{2}K\mu(E_x+E_y)$.

The experimental confirmation of the field-mediated spin effects
predicted here would stimulate further progress both in basic
research and technological innovations.

\begin{acknowledgments}
Partial financial support by the Deutsche Forschungsgemeinschaft
and the Russian Foundation of Basic Research is gratefully
acknowledged.
\end{acknowledgments}
\appendix
\section{Fourier transformed drift-diffusion equations}
The Fourier transformed version of our basic spin-charge coupled
drift-diffusion equations (\ref{b1}) and (\ref{JJ}) has the form
\begin{equation}
\left[\frac{\partial}{\partial t}-i\mu{\bm{E}}{\bm{\kappa}}
+D\kappa^2 \right]F+i{\bm{\omega}}_{\bm{\kappa}}\cdot{\bm{F}}
-\frac{2im\tau}{\hbar^3}(\alpha^2-\beta^2)([{\bm{\kappa}}\times
\mu{\bm{E}}]\cdot {\bm{F}})=0,\label{A1}
\end{equation}
\begin{equation}
\left[\frac{\partial}{\partial t}-i\mu{\bm{E}}{\bm{\kappa}}
+D\kappa^2+\frac{\overleftrightarrow{T}}{\tau_s}
\right]\bm{F}-\left[ \bm{H}\times\bm{F}\right]-
\frac{i\chi}{\hbar}\left[\bm{\kappa}\times\mu\bm{E}
\right]F+\chi\bm{H}F=0,\label{A2}
\end{equation}
with the abbreviations
\begin{equation}
H_x=\frac{2m}{\hbar^2}\left[\alpha(\mu E_y+2iD\kappa_y)+\beta(\mu
E_x+2iD\kappa_x) \right],
\end{equation}
\begin{equation}
H_y=-\frac{2m}{\hbar^2}\left[\alpha(\mu E_x+2iD\kappa_x)+\beta(\mu
E_y+2iD\kappa_y) \right],\quad H_z=0,
\end{equation}
\begin{equation}
\overleftrightarrow{T}=\left(%
\begin{array}{ccc}
  1 & \frac{2\alpha\beta}{\alpha^2+\beta^2} & 0 \\
  \frac{2\alpha\beta}{\alpha^2+\beta^2} & 1 & 0 \\
  0 & 0 & 2 \\
\end{array}%
\right),\quad \chi=\frac{2m\tau}{\hbar^3}(\alpha^2-\beta^2).
\end{equation}
For the Rashba model ($\beta=0$), these equations agree with
results derived previously.~\cite{PRB_075340}

Let us treat these equations in another representation
characterized by the components
\begin{equation}
\kappa_{\pm}=\frac{\kappa_x \pm \kappa_y}{\sqrt{2}},\quad
F_{\pm}=\frac{F_x\pm F_y}{\sqrt{2}}.
\end{equation}
For excitations of the spin polarization that propagate
exclusively along the $\kappa_{+}$ direction ($\kappa_{-}=0$),
Eqs. (\ref{A1}) and (\ref{A2}) are written in the form
\begin{equation}
\left[\frac{\partial}{\partial t}-i\mu
E_{+}\kappa_{+}+D\kappa_{+}^2 \right]F-\frac{i}{\hbar}(\alpha
+\beta)\kappa_{+}F_{-}+\frac{2im\tau}{\hbar^3}(\alpha^2-\beta^2)\mu
E_{-}\kappa_{+}F_z=0,\label{AA1}
\end{equation}
\begin{eqnarray}
&&\left[\frac{\partial}{\partial t}-i\mu
E_{+}\kappa_{+}+D\kappa_{+}^2+\frac{2}{\tau_{s+}}
\right]F_{+}+\frac{2m}{\hbar^2}(\alpha+\beta)(\mu
E_{+}+2iD\kappa_{+})F_z\nonumber\\
&&-\frac{4m^2\tau}{\hbar^5}(\alpha-\beta)^2(\alpha+\beta)\mu
E_{-}F=0,\label{AA2}
\end{eqnarray}
\begin{eqnarray}
&&\left[\frac{\partial}{\partial t}-i\mu
E_{+}\kappa_{+}+D\kappa_{+}^2+\frac{2}{\tau_{s-}}
\right]F_{-}+\frac{2m}{\hbar^2}(\alpha-\beta)\mu
E_{-}F_z\nonumber\\
&&+\frac{4m^2\tau}{\hbar^5}(\alpha+\beta)^2(\alpha-\beta)(\mu
E_{+}+2iD\kappa_{+})F=0,\label{AA3}
\end{eqnarray}
\begin{eqnarray}
&&\left[\frac{\partial}{\partial t}-i\mu
E_{+}\kappa_{+}+D\kappa_{+}^2+\frac{2}{\tau_s}
\right]F_{z}+\frac{2im\tau}{\hbar^3}(\alpha^2-\beta^2)\mu
E_{-}\kappa_{+}F\nonumber\\
&&-\frac{2m}{\hbar^2}(\alpha+\beta)(\mu
E_{+}+2iD\kappa_{+})F_{+}-\frac{2m}{\hbar^2}(\alpha-\beta)\mu
E_{-}F_{-}=0,\label{AA4}
\end{eqnarray}
with $E_{\pm}=(E_x\pm E_y)/\sqrt{2}$ and
\begin{equation}
\frac{2}{\tau_{s+}}=\frac{(\alpha+\beta)^2}{\tau_s(\alpha^2+\beta^2)},\quad
\frac{2}{\tau_{s-}}=\frac{(\alpha-\beta)^2}{\tau_s(\alpha^2+\beta^2)}.
\end{equation}
For the particular case $\alpha=\beta$, the coupling between spin
and charge degrees of freedom disappears and the secular equation
for the spin components gives the following dispersion relations
for eigenmodes of the biased spin system
\begin{equation}
\omega_1=-\mu E_{+}\kappa_{+}-iD\kappa_{+}^2,\quad
\omega_{2,3}=-(\kappa_{+}\pm 2K)\left[\mu E_{+}+iD(\kappa_{+}\pm
2K) \right],
\end{equation}
with $K=2m\alpha/\hbar^2$. This result implies that there appears
a field-induced undamped soft mode at $\kappa_{+}=2K$, which
reflects the presence of the spin rotation
symmetry.~\cite{PRL_236601}


\begin{thebibliography}{10}

\bibitem{Dyakonov}
M.~I. Dyakonov and V.~I. Perel, JETP Lett. {\bf 13}, 467, (1971),
[Sov. Phys. JETP 33, 1053 (1971)].

\bibitem{JAP_073702}
J.~L. Cheng and M.~W. Wu, J. Appl. Phys. {\bf 101}, 073702 (2007).

\bibitem{PRB_15582}
N.~S. Averkiev and L.~E. Golub, Phys. Rev. B {\bf 60}, 15582
(1999).

\bibitem{JPC_R271}
N.~S. Averkiev, L.~E. Golub, and M. Willander, J. Phys.: Condens.
Matter {\bf 14}, R271 (2002).

\bibitem{PRL_146801}
J. Schliemann, J.~C. Egues, and D. Loss, Phys. Rev. Lett. {\bf
90}, 146801 (2003).

\bibitem{PRB_165311}
J. Schliemann and D. Loss, Phys. Rev. B {\bf 68}, 165311 (2003).

\bibitem{Pershin_E}
Y.~V. Pershin, Physica E {\bf 23}, 226 (2004).

\bibitem{PRB_081312}
N.~A. Sinitsyn, E.~M. Hankiewicz, W. Teizer, and J. Sinova, Phys.
Rev. B {\bf 70}, 081312 (2004).

\bibitem{PRB_085315}
P.~Q. Jin and Y.~Q. Li, Phys. Rev. B {\bf 74}, 085315 (2006).

\bibitem{PRB_121308}
A.~G. Mal'shukov and K.~A. Chao, Phys. Rev. B {\bf 71}, 121308
(2005).

\bibitem{PRB_085344}
X.~F. Wang and P. Vasilopoulos, Phys. Rev. B {\bf 72}, 085344
(2005).

\bibitem{IJMPB_4937}
V.~V. Bryksin and P. Kleinert, Int. J. Mod. Phys. B {\bf 20}, 4937
(2006).

\bibitem{PRB_155323}
M. Trushin and J. Schliemann, Phys. Rev. B {\bf 75}, 155323
(2007).

\bibitem{PRL_236601}
B.~A. Bernevig, J. Orenstein, and S.~C. Zhang, Phys. Rev. Lett.
{\bf 97}, 236601 (2006).

\bibitem{PRB_235322}
M.~H. Liu, K.~W. Chen, S.~H. Chen, and C.~R. Chang, Phys. Rev. B
{\bf 74}, 235322 (2006).

\bibitem{PRL_076604W}
C.~P. Weber, J. Orenstein, B.~A. Bernevig, S.~C. Zhang, J.
Stephens, and D.~D. Awschalom, Phys. Rev. Lett. {\bf 98}, 076604
(2007).

\bibitem{PRB_241308}
M. Ohno and K. Yoh, Phys. Rev. B {\bf 75}, 241308 (2007).

\bibitem{PRB_125307}
T.~D. Stanescu and V. Galitski, Phys. Rev. B {\bf 75}, 125307
(2007).

\bibitem{SSC_139}
P. Kleinert and V. Bryksin, Solid State Commun. {\bf 139}, 205
(2006).

\bibitem{PRB_165313}
V.~V. Bryksin and P. Kleinert, Phys. Rev. B {\bf 73}, 165313
(2006).

\bibitem{PRB_075340}
V.~V. Bryksin and P. Kleinert, Phys. Rev. B {\bf 76}, 075340
(2007).

\bibitem{PRB_041308}
V.~L. Korenev, Phys. Rev. B {\bf 74}, 041308 (2006).

\bibitem{ssc_559}
V.~K. Kalevich, V.~L. Korenev, and I.~A. Merkulov, Solid State
Commun. {\bf 91}, 559 (1994).

\bibitem{PRB_235302}
V.~V. Bryksin, H. B\"ottger, and P. Kleinert, Phys. Rev. B {\bf
74}, 235302 (2006).

\bibitem{Kalevich}
V.~K. Kalevich and V.~L. Korenev, JETP Lett. {\bf 52}, 230 (1990),
[Pis'ma Zh. Eksp. Teor. Fiz. {\bf 52}, 859 (1990)].

\bibitem{PRB_205327}
T. Damker, H. B\"ottger, and V.~V. Bryksin, Phys. Rev. B {\bf 69},
205327 (2004).

\bibitem{PRB_205202}
O. Bleibaum, Phys. Rev. B {\bf 69}, 205202 (2004).

\bibitem{PRB_235318}
O. Bleibaum, Phys. Rev. B {\bf 71}, 235318 (2005).

\bibitem{PRL_4196}
Y. Ohno, R. Terauchi, T. Adachi, F. Matsukura, and H. Ohno, Phys.
Rev. Lett. {\bf 83}, 4196 (1999).

\bibitem{PRL_036603}
O.~D.~D. Couto, F. Iikawa, J. Rudolph, R. Hey, and P.~V. Santos,
Phys. Rev. Lett. {\bf 98}, 036603 (2007).

\end{thebibliography}


\end{document}